\documentclass[a4paper,12pt]{article}
\usepackage[top=2.8cm,bottom=2.8cm,left=1.8cm,right=1.8cm]{geometry}
\usepackage{graphicx}
\usepackage{amsmath}
\usepackage{amsthm}
\usepackage{amsfonts}
\usepackage{amssymb}
\usepackage{mathrsfs}
\usepackage{caption}
\usepackage{subcaption}
\usepackage{hyperref}
\newcommand\bb[1]{{\boldsymbol{#1}}}
\newcommand\der[2]{\frac{{\rm{d}}#1}{{\rm{d}}#2}}
\providecommand{\keywords}[1]{\textbf{Keywords:} #1}
\DeclareMathOperator{\dist}{dist}
\title{An Edge-Based Interface-Tracking Method for Multiphase Flows}

\author{Leonardo Chirco$^1$ \\ \small  leonardo.chirco@sorbonne-universite.fr 
   \and \large St\'ephane Zaleski $^{1,2}$ \\ \small stephane.zaleski@sorbonne-universite.fr }
\date{%
\small$^1$Sorbonne Universit\'e and CNRS, Institut Jean Le Rond d'Alembert \\
UMR 7190, F-75005 Paris, France\\%
    $^2$Institut Universitaire de France, Paris, France\\[2ex]%
    \today
}
\begin{document}
\maketitle
\begin{abstract}
\noindent
 We propose a novel class of Edge-Based Interface-Tracking (EBIT) methods 
in the field of multiphase flows for advecting the interface. 
The position of the interface is tracked by marker points located on 
the edges of the underlying grid, making the method flexible
with respect to the choice of spatial discretization and suitable for parallel computation.
In this paper we present a simple EBIT method based on two-dimensional Cartesian grids
and on a linear interface representation.
\end{abstract}
%%%%%%%%%%%%%%%%%%%%%%%%%%%%%%%%%%%%%%%%%%%%%%%%%%%%%%%%%%%%%%%%%%%%%%%%%%%%%%%
\section{Introduction}
Many methods for following an interface or front exist, 
the simplest and most popular being the Front-Tracking,
the Level-Set and the Volume-of-Fluid method \cite{tryggvason2011}.
In this paper we first consider a new class of methods,
which could be called Edge-Based Interface-Tracking (EBIT) methods.
In these methods, the basic information about the front position is known or ``tracked''
by the position of marker points, which makes the method a kind of Front-Tracking.
However, the additional requirement is that the markers
are located on the edges of the underlying grid. 
When the connecting interface lines between the marker points are linear,
the method bears an obvious similarity with 
the Volume-Of-Fluid method of Piecewise Linear Interface Calculation type (PLIC-VOF).
Finally, since the position of the markers gives an explicit information
about the distance of the vertices of the underlying grid to the interface, 
it is a kind of distance information as in the Level-Set method,
where the implicit definition of the interface is given by a function
as close as possible to the signed distance function. In particular,
a linear interface has the same representation using EBIT and Level-Set methods.

\begin{figure}
\centering
    \includegraphics[width=0.95\textwidth]{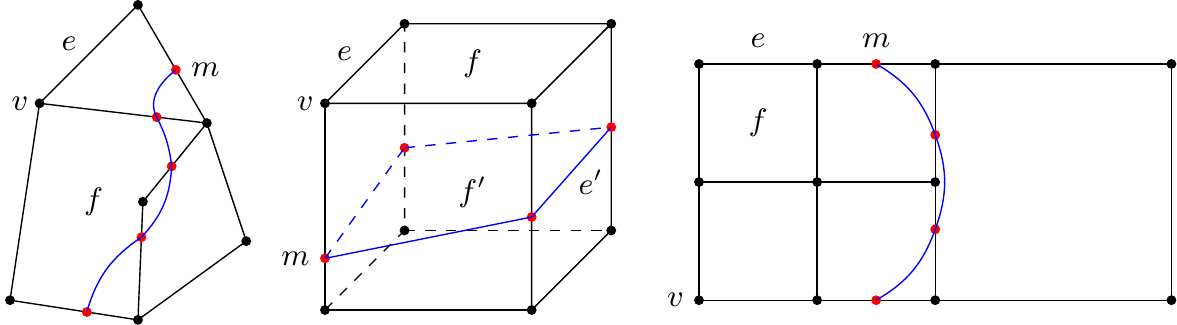}
    \caption{Schematics of Edge-Based Interface-Tracking. 
    (Left) On an unstructured planar grid formed of edges $e$ (black lines),
    vertices $v$ (black dots) and faces $f$ (polygons delimited by black lines),
    the interface passes through the markers $m$ (red dots). 
    (Center) On a regular cuboid volume grid, with again edges $e$ (black lines),
    vertices $v$ (black dots) and faces $f$ (squares delimited by black lines),
    the interface passes  through the markers $m$ (red dots).
    The markers $m$ form the vertices of the surface
    grid tracking the interface, with edges $e'$ and faces $f'$ on the latter grid.
    The faces are in general non-planar. 
    (Right) The leaf cells on a quadtree grid form a particular type of unstructured grid.
    The edges can again be the location of marker points.}
    \label{imm:ebit}
\end{figure}

Several prior works have attempted a combination of pairs of the three main methods
and may result in methods similar to this one, 
such as the combination of markers and VOF \cite{aulisa2004surface}
or the combination of Level-Set and Front-Tracking \cite{shin2009hybrid}.
However, the EBIT method adds the simplifying requirement that
only the position of the markers on the grid lines or grid edges needs to be known.
This is true both in 2D and 3D and whatever the grid type, structured,
unstructured or hierarchical/quadtree, see Figure \ref{imm:ebit}.
The use of iso-faces to perform the advection of interfaces 
on general meshes consisting of arbitrary polyhedral cell
is the core of the \textit{isoAdvector} algorithm as well, see \cite{isoadvector}. 
Perhaps the most important advantage of EBIT methods is 
that they allow for almost automatic parallelization.
In fact, since the marker points are constrained to move along the grid edges,
their re-distribution among processes 
follows naturally that of the grid cells.
Another potential advantage is that as the grid is adapted,
refined or unrefined the front is adapted consistently.
Finally, since information about the connectivity of the marker points does not need to be stored
(it can be reconstructed and is thus known implicitly)
the addition or removal of points or grid cells is  easier than in traditional Front-Tracking \cite{glimm2000robust}.
In this paper we focus on a special case of EBIT methods, the Semushin method,
in which the underlying grid is a 2D square grid, 
the intersections are at most two per square edge of the grid
and the interpolation between the marker points is linear.
This is clearly a ``bare bones'' version of the EBIT method and is inspired by Semushin's preprint \cite{sem}
and by personal communications received from its author.
This article is organized as follows. The method is described is Section \ref{sec:met}
and then in Section \ref{sec:resu} the numerical results are presented.
Finally, the conclusions are given in the last section.
%%%%%%%%%%%%%%%%%%%%%%%%%%%%%%%%%%%%%%%%%%%%%%%%%%%%%%%%%%
%%%%%%%%%%%%%%%%%%%%%%%%%%%%%%%%%%%%%%%%%%%%%%%%%%%%%%%%%%
%%%%%%%%%%%%%%%%%%%%%%%%%%%%%%%%%%%%%%%%%%%%%%%%%%%%%%%%%%%%%%%%%%%%%%%%%%%%%%%
\section{The Semushin method} \label{sec:met}
In Semushin's method for tracking the interface, the reference phase
is enclosed by a set of marker points placed on the grid lines.
The advection of the interface is done by moving these points along the grid lines.
Thanks to this constraint, the $n$-dimensional advection algorithm can be split into 
a succession of $n$ times the one-dimensional scheme, one for each direction. 
%%%%%%%%%%%%%%%%%%%%%%%%%%%%%%%%%%%%%%%%%%%%%%%%%%%%%%%%%%
\begin{figure}[htb!]
 \centering
 \begin{subfigure}[t]{0.24\textwidth}
    \includegraphics[width=\textwidth]{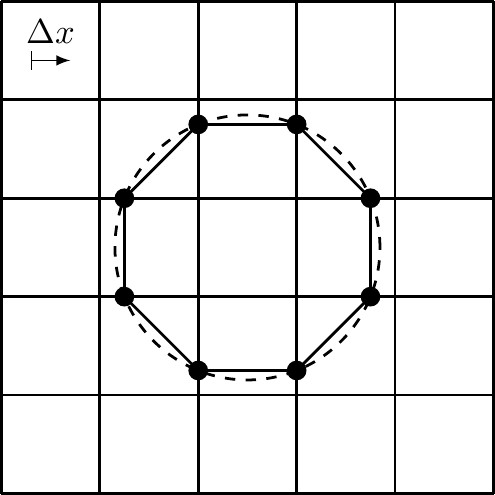}
    \caption{Initial markers position for the dashed circle.}
    \label{imm:sem-1}
 \end{subfigure}
 \hfill
 \begin{subfigure}[t]{0.24\textwidth}
    \includegraphics[width=\textwidth]{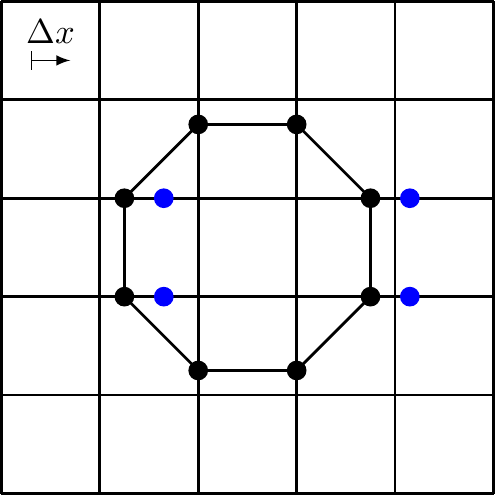}
    \caption{Advection of the (blue) points aligned with the velocity.}% The blue points are in their final position.}
    \label{imm:sem-2}
 \end{subfigure}
 \hfill
 \begin{subfigure}[t]{0.24\textwidth}
    \includegraphics[width=\textwidth]{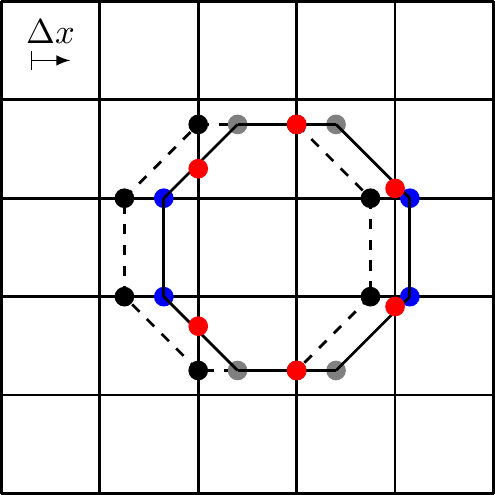}
    \caption{Fictitious advection of the unaligned (gray) points and (red) intersections.}
    \label{imm:sem-3}
 \end{subfigure}
  \hfill
 \begin{subfigure}[t]{0.24\textwidth}
    \includegraphics[width=\textwidth]{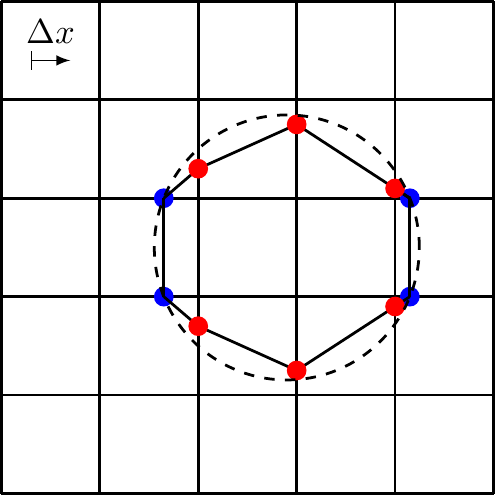}
    \caption{Final position of the markers and interface.}
    \label{imm:sem-4}
 \end{subfigure}
     \caption{The steps of the one-dimensional advection scheme of Semushin's method.}
    \label{imm:sem}
\end{figure}
%%%%%%%%%%%%%%%%%%%%%%%%%%%%%%%%%%%%%%%%%%%%%%%%%%%%%%%%%%
The equation of motion for the interface point is
%%%%%%%%%%%%%%%%%%%%%%%%%%%%%%%%%%%%%%%%%%%%%%%%%%%%%%%%%%
\begin{equation}
 \der{\bb{x}}{t}=\bb{u} \,,
\end{equation}
%%%%%%%%%%%%%%%%%%%%%%%%%%%%%%%%%%%%%%%%%%%%%%%%%%%%%%%%%%
that can be integrated as
%%%%%%%%%%%%%%%%%%%%%%%%%%%%%%%%%%%%%%%%%%%%%%%%%%%%%%%%%%
\begin{equation} \label{eq:mot}
  \bb{x}=\bb{x}_0+\int_{t_0}^t u(\bb{x}(t'),t')\rm{d}t'\,,
\end{equation}
%%%%%%%%%%%%%%%%%%%%%%%%%%%%%%%%%%%%%%%%%%%%%%%%%%%%%%%%%%
where the initial position $\bb{x}_0$ is known.
For the sake of simplicity, in this work we use a first-order explicit Euler method
such that $x=x_0+u_0\Delta t=x_0+\Delta x$.

Now, we describe the simple one-dimensional advection algorithm used, see Figure \ref{imm:sem}.
We recall that we study a two-dimensional problem,
admit at most two interface intersections (and then markers)
per face (edge in 2D) of the grid, and that the interpolation between the marker points is linear.
The extension to three-dimensional problems or unstructured grids is straightforward,
see Figure \ref{imm:ebit}.
The points placed on the grid lines aligned with the velocity are called \textit{aligned} points,
while the remaining ones are \textit{unaligned}.
Starting from the initial configuration (Figure \ref{imm:sem-1}), 
the new position of the aligned points (Figure \ref{imm:sem-2}) is directly obtained by integrating \eqref{eq:mot}.
To place the unaligned points (in this example on the vertical grid lines),
we first advect them using the same equation \eqref{eq:mot} obtaining the \textit{fictitious}
gray points in Figure \ref{imm:sem-3}.
Finally, the new position of the unaligned points (in red in Figure \ref{imm:sem-3}) 
is obtained by connecting with a segment either one blue and one gray point 
or two consecutive gray points and by finding the intersection with the grid lines.
The position of the points and of the interface after the advection
along the $x$-direction is shown in Figure \ref{imm:sem-4}.

%%%%%%%%%%%%%%%%%%%%%%%%%%%%%%%%%%%%%%%%%%%%%%%%%%%%%%%%%%%%%%%%%%%%%%%%%%%%%%%
\section{Results}
We define the surface error $E_{area}(t)$ between the
total area of the reference phase at the initial time $t_0$
and time $t$ as
%%%%%%%%%%%%%%%%%%%%%%%%%%%%%%%%%%%%%%%%%%%%%%%%%%%%%%%%%%
\begin{equation}  \label{eq:m_err}
 E_{area}=\frac{|A(t)-A(t_0)|}{A(t_0)} \,.
\end{equation}
%%%%%%%%%%%%%%%%%%%%%%%%%%%%%%%%%%%%%%%%%%%%%%%%%%%%%%%%%%
We define the shape error, in a $L^\infty$ norm, as the maximum
distance between any marker point $\bb{x_i}$ on the interface and
the corresponding closest point on the analytical shape
as
%%%%%%%%%%%%%%%%%%%%%%%%%%%%%%%%%%%%%%%%%%%%%%%%%%%%%%%%%%
\begin{equation}  \label{eq:l_inf_err}
 E_{shape}=\max_{i}|\dist(\bb{x}_i)| \,.
\end{equation}
%%%%%%%%%%%%%%%%%%%%%%%%%%%%%%%%%%%%%%%%%%%%%%%%%%%%%%%%%%
We recall that for a circle centered in $(x_c,y_c)$ and radius $R$, we have 
$\dist(\bb{x}_i)=\sqrt{(x_i-x_c)^2+(y_i-y_c)^2}-R$.
The order of convergence of the method is computed by comparing the errors 
on successively refined grids as
%%%%%%%%%%%%%%%%%%%%%%%%%%%%%%%%%%%%%%%%%%%%%%%%%%%%%%%%%%
\begin{equation} \label{eq:order}
 \mathrm{order}=\log_2(E(h)/E(h/2)) \,,
\end{equation}
%%%%%%%%%%%%%%%%%%%%%%%%%%%%%%%%%%%%%%%%%%%%%%%%%%%%%%%%%%
where $E(h)$ is the norm of the error on the grid with spacing $h$,
with respect to the exact solution.
We perform three well-known tests to evaluate the accuracy of interface advecting
methods \cite{aulisa2003geometrical}. 

\paragraph{Translation with uniform velocity}
%%%%%%%%%%%%%%%%%%%%%%%%%%%%%%%%%%%%%%%%%%%%%%%%%%%%%%%%%%%
\begin{figure}[htb!]
 \centering
  \begin{subfigure}[t]{0.4\textwidth}
    \includegraphics[width=\textwidth]{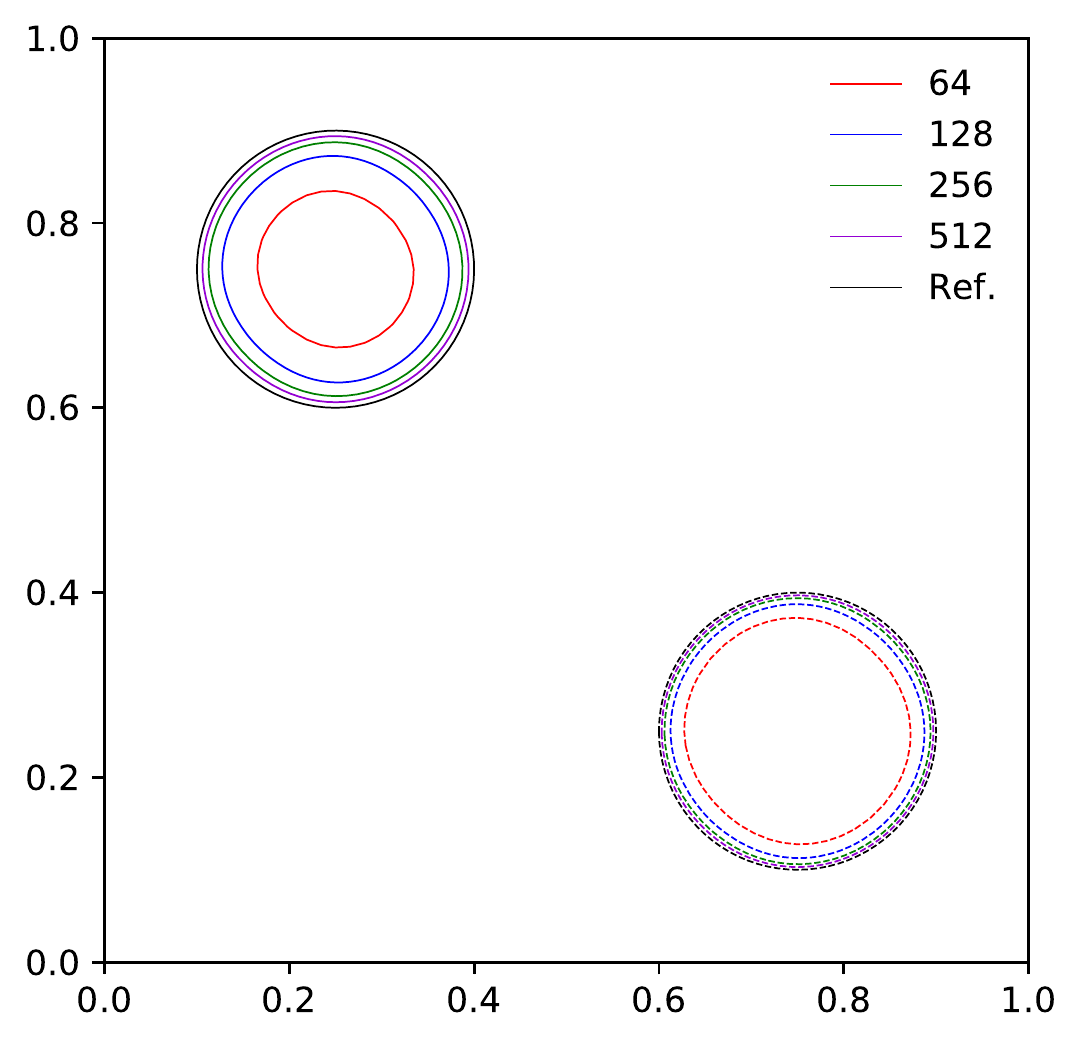}
    \caption{$\textit{CFL}=0.125$.}
    \label{imm:CFL0p125}
 \end{subfigure}
  \begin{subfigure}[t]{0.4\textwidth}
    \includegraphics[width=\textwidth]{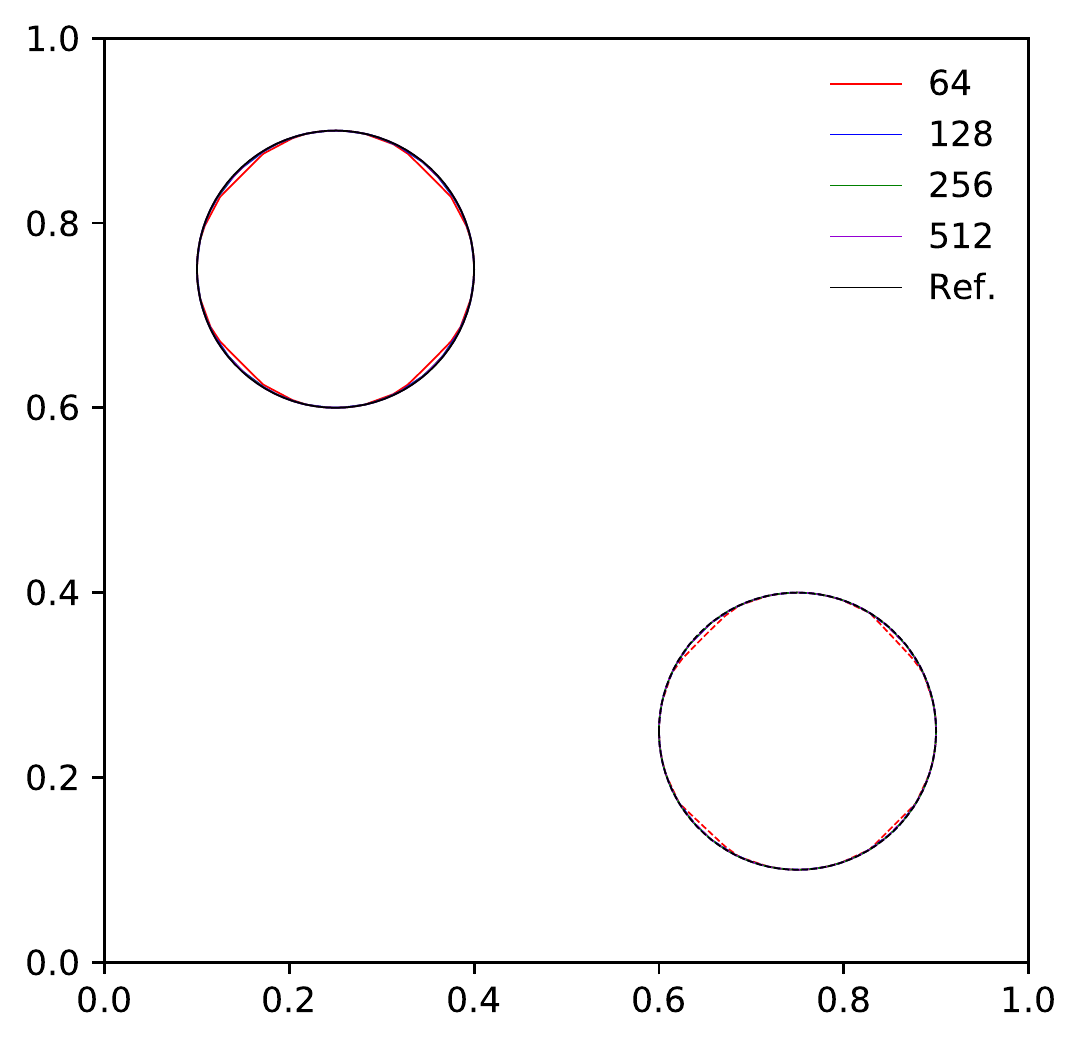}
    \caption{$\textit{CFL}=1$.}
    \label{imm:CFL1}
 \end{subfigure}
 \caption{Final circular shape (solid line) and after half diagonal translation (dashed line).}
 \label{imm:transla}
\end{figure}
 %%%%%%%%%%%%%%%%%%%%%%%%%%%%%%%%%%%%%%%%%%%%%%%%%%%%%%%%%%%
%%%%%%%%%%%%%%%%%%%%%%%%%%%%%%%%%%%%%%%%%%%%%%%%%%%%%%%%%%%
\begin{table}[htb!]
\centering
   \begin{tabular}{ccccc} 
     \hline
     $\mathit{N_x}$ & \textit{CFL} & $\mathit{E_{area}}$ & $\mathit{E_{shape}}$ & Order \\
      \hline
      $64  $ &$1.0  $ & $2.89e-2$ & $2.78e-2$ & $2.17$ \\
             &$0.125$ & $6.89e-1$ & $7.01e-2$ & $1.01$ \\
      &&&\\
      $128 $ &$1.0  $ & $6.42e-3$ & $1.23e-2$ & $1.33$ \\
             &$0.125$ & $3.43e-1$ & $3.12e-2$ & $1.00$ \\      
      &&&\\
      $256 $ &$1.0  $ & $2.56e-3$ & $6.33e-3$ &$0.93$ \\
             &$0.125$ & $1.72e-1$ & $1.55e-2$ &$1.00$ \\      
      &&&\\
      $512 $ &$1.0  $ & $1.34e-3$ & $3.57e-3$ &\\
             &$0.125$ & $8.57e-2$ & $7.87e-3$ &\\      
      \hline
  \end{tabular}
  \caption{Surface error $\mathit{E_{area}}$, shape error $\mathit{E_{shape}}$, and order of convergence for two complete translations along the main diagonal, at different resolutions and \textit{CFL} numbers.}
  \label{tab:transla}
\end{table}
%%%%%%%%%%%%%%%%%%%%%%%%%%%%%%%%%%%%%%%%%%%%%%%%%%%%%%%%%%%
In the first test a circular shape of radius $r=0.15$
and center $(0.25,0.75)$ is placed inside the unit box.
The box is meshed with $N_x\times N_x$ square cells of size $h=1/N_x$, 
where $N_x=64,128,256,512$. 
A uniform and constant velocity field $(u,v)$ with $u=-v$ is imposed in the box,
so that the reference phase is advected along the diagonal of the box.
After one time unit, the velocity field is reversed and the circular fluid body
should return to its initial position with no distortion,
allowing error measurement with \eqref{eq:m_err} and \eqref{eq:l_inf_err}.
For this test, we employ two constant \textit{CFL} numbers $\textit{CFL}=u\Delta t/h $,
where $\Delta t$ is the time step. For example, if $\textit{CFL}=1$, 
the circle is displaced of exactly one grid spacing per time step, while if $\textit{CFL}<1$,
the circle advances only by a fraction of the grid spacing.

\begin{figure}[htb!]
 \centering
   \includegraphics[width=0.4\textwidth]{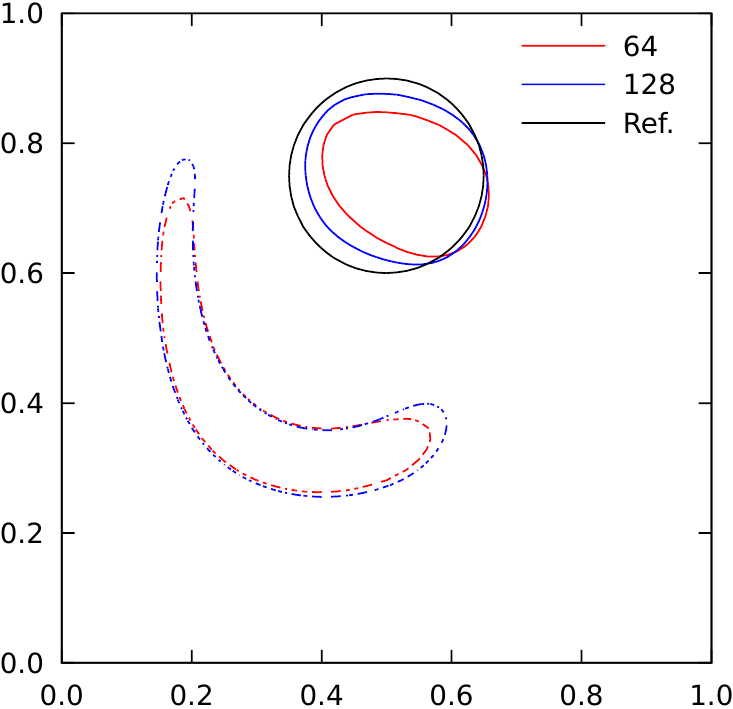}
   \includegraphics[width=0.4\textwidth]{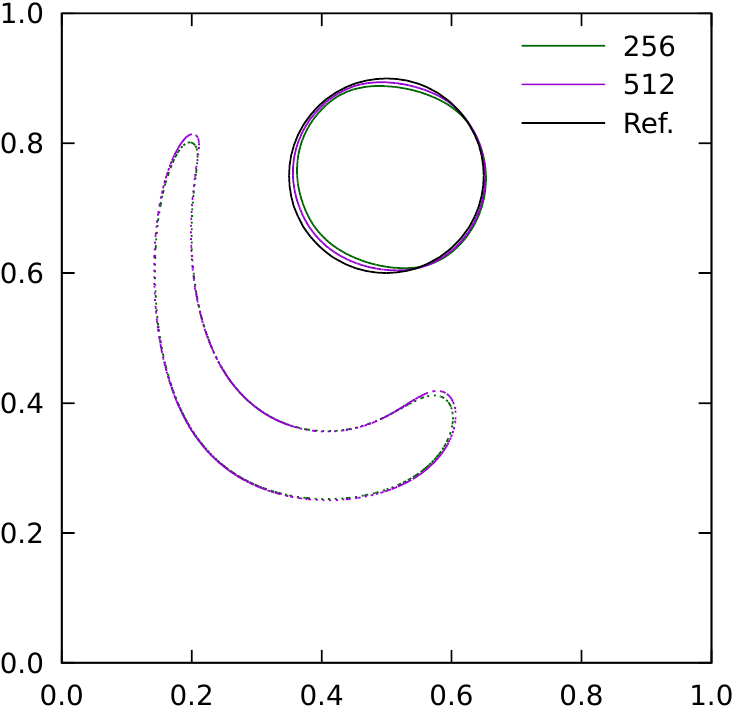}
   \caption{
   The interface at maximum deformation at $t=1.0$ (dotted line)
   and back to the initial position at $t=2.0$ (solid line) for the single vortex field test with $T=2.0$.}
   \label{imm:vortex}
\end{figure}

In Figure \ref{imm:transla}, the position of the reference phase is shown
after two full diagonal translations (solid line) and one (dashed line).
When using the coarser grids, the circular shape is shrunk radially.
In Table \ref{tab:transla} we report the surface error $\mathit{E_{area}}$,
the shape error $\mathit{E_{shape}}$,
and order of convergence for two complete translations along the main diagonal,
at different resolutions and \textit{CFL} numbers.
In purely kinematic tests, smaller errors are obtained using
$\textit{CFL} = 1$, since fewer substeps of the algorithm are 
necessary to obtain a given displacement. 
However, since the intended use of EBIT methods is advecting the interface
in multiphase flows where the \textit{CFL} 
has to be limited for stability reasons, the accumulation of errors 
will affect the performance.
%%%%%%%%%%%%%%%%%%%%%%%%%%%%%%%%%%%%%%%%%%%%%%%%%%%%%%%%%%%%%%%%%%%%%%%%%%%%%%%
\paragraph{Single vortex rotation} \label{sec:resu}
The single vortex or ``vortex-in-a-box'' problem has been designed to test
the ability of interface tracking methods
when the reference phase is highly stretched, see \cite{enright2002hybrid}.
A circular shape of radius $r=0.15$
and center $(0.5,0.75)$ is placed inside the unit box.
The divergence-free velocity $\bb{u} = (u , v)$
is obtained from the following stream function
%%%%%%%%%%%%%%%%%%%%%%%%%%%%%%%%%%%%%%%%%%%%%%%%%%%%%%%%%%
% \begin{equation}
 $\psi = \pi^{-1} \sin^2 (\pi x) \sin^2 (\pi y) \cos(\pi t/T) \,,$
% \end{equation}
%%%%%%%%%%%%%%%%%%%%%%%%%%%%%%%%%%%%%%%%%%%%%%%%%%%%%%%%%%
as $u_x = \partial \psi/\partial y$ and $u_y = -\partial \psi/\partial x$.
On the sides of the box, homogeneous Dirichlet boundary conditions are imposed.
The cosinusoidal time-dependence slows down and reverses the flow, so that
the maximum deformation occurs at $t=T/2$ and at time
$T$ the reference phase returns to its initial position with no distortion,
allowing again to measure the error with \eqref{eq:m_err} and \eqref{eq:l_inf_err}, 
see \cite{leveque1996}.
For this test we use a constant time step $\Delta t=0.0005$.
The position of the reference phase at $t=1$, corresponding to its maximum deformation,
and at $t=T=2$ back to the initial position is shown in Figure \ref{imm:vortex}.
By refining the grid, the main fluid becomes thinner and more elongated at $t=1$,
while tends to the reference initial shape at $t=2$.
In Table \ref{tab:vortex} we report the surface error $\mathit{E_{area}}$,
the shape error $\mathit{E_{shape}}$,
and order of convergence at different grid resolutions.
%%%%%%%%%%%%%%%%%%%%%%%%%%%%%%%%%%%%%%%%%%%%%%%%%%%%%%%%%%%
 %%%%%%%%%%%%%%%%%%%%%%%%%%%%%%%%%%%%%%%%%%%%%%%%%%%%%%%%%%%
%%%%%%%%%%%%%%%%%%%%%%%%%%%%%%%%%%%%%%%%%%%%%%%%%%%%%%%%%%%
\begin{table}[htb!]
\centering
   \begin{tabular}{cccc} 
     \hline
      $\mathit{N_x}$ & $\mathit{E_{area}}$ & $\mathit{E_{shape}}$  & Order \\
      \hline
      $64  $ & $3.77e-1$ & $6.43e-2$ & $1.14$ \\
      &&&\\
      $128 $ & $1.71e-1$ & $2.95e-2$ & $1.09$ \\
      &&&\\
      $256 $ & $8.04e-2$ & $1.45e-2$ & $1.10$ \\
      &&&\\
      $512 $ & $3.76e-2$ & $7.42e-3$\\
      \hline
  \end{tabular}
  \caption{Surface error $\mathit{E_{area}}$, shape error $\mathit{E_{shape}}$, and order of convergence for the single vortex test with $T=2.0$, at different resolutions.}
  \label{tab:vortex}
\end{table}
%%%%%%%%%%%%%%%%%%%%%%%%%%%%%%%%%%%%%%%%%%%%%%%%%%%%%%%%%%%

%%%%%%%%%%%%%%%%%%%%%%%%%%%%%%%%%%%%%%%%%%%%%%%%%%%%%%%%%%%%%%%%%%%%%%%%%%%%%%%
\paragraph{Zalesak’s disk rotation} 
%%%%%%%%%%%%%%%%%%%%%%%%%%%%%%%%%%%%%%%%%%%%%%%%%%%%%%%%%%%
\begin{figure}[htb!]
 \centering
      \includegraphics[width=0.34\textwidth]{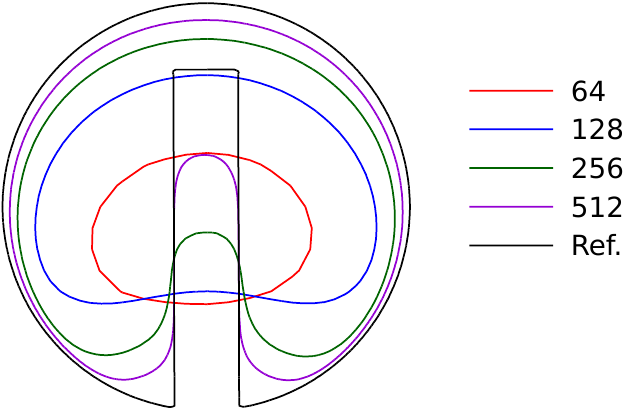}
   \caption{Initial (Ref.) and final shape for the Zalesak’s disk after one rotation at $T=1.0$.}
   \label{imm:zalesak}
\end{figure}
%%%%%%%%%%%%%%%%%%%%%%%%%%%%%%%%%%%%%%%%%%%%%%%%%%%%%%%%%%%
%%%%%%%%%%%%%%%%%%%%%%%%%%%%%%%%%%%%%%%%%%%%%%%%%%%%%%%%%%%
\begin{table}[htb!]
\centering
   \begin{tabular}{ccc} 
     \hline
      $\mathit{N_x}$ & $\mathit{E_{area}}$ & Order \\
      \hline
      $64  $ & $7.56e-1$ & $0.87$ \\
      &&\\
      $128 $ & $4.12e-1$ & $1.27$ \\
      &&\\
      $256 $ & $1.71e-1$ & $1.36$ \\
      &&\\
      $512 $ & $6.69e-2$ &\\
      \hline
  \end{tabular}
  \caption{Surface error $\mathit{E_{area}}$ and order of convergence for the Zalesak's disk rotation test with $T=1.0$, at different resolutions.}
  \label{tab:zalesak}
\end{table}
%%%%%%%%%%%%%%%%%%%%%%%%%%%%%%%%%%%%%%%%%%%%%%%%%%%%%%%%%%%
In this test a notched circle of radius $r=0.15$ and center $(0.5,0.75)$
is placed inside the unit box. The notched width is $0.05$ 
and the length is $0.25$. Imposing the constant velocity field
$(u,v)= (2\pi(0.5-y), 2\pi(x-0.5))$ the disk performs a full rotation around
the box center and returns to the initial position at $T=1.0$. %CIT if enough space!
At the lowest resolution the notch disappears, while increasing the resolution
the notch is maintained with smoothed corners. Interestingly, our method 
recovers final shapes that are symmetrical with respect to the notch vertical axis,
which is not always observed in literature especially at low resolution, see
\cite{henri2022, boniou2022}. 
In Table \ref{tab:zalesak} we report the surface error $\mathit{E_{area}}$
and the order of convergence at different grid resolutions.
The method exhibits a first-order convergence rate upon grid refinement.

%%%%%%%%%%%%%%%%%%%%%%%%%%%%%%%%%%%%%%%%%%%%%%%%%%%%%%%%%%%%%%%%%%%%%%%%%%%%%%%
\section{Conclusions}
In this article, we have studied a new Interface-Tracking method, where
the interface is tracked by marker points located on 
the edges of the underlying grid.
We have implemented the two-dimensional version of the method
using linear interface reconstruction.
We have used three well-known benchmark tests to validate the numerical method,
recovering a first-order convergence rate of the surface error, lower than
the one obtained with other methods, such as VOF, Level-Set or isoAdvector, 
\cite{henri2022,boniou2022,isoadvector}.
In future works we aim to use the EBIT method for multiphase simulations,
developing models for topology changes and surface tension and extending the
method to three dimensions.
%%%%%%%%%%%%%%%%%%%%%%%%%%%%%%%%%%%%%%%%%%%%%%%%%%%%%%%%%%%%%%%%%%%%%%%%%%%%%%%
\subsection{Acknowledgements} 
St\'ephane Zaleski recalls meeting Sergei Semushin in March 1995 and learning about his method. He thanks him for the explanation of the method and the gift of the preprint \cite{sem}.
The authors benefited from the ERC grant TRUFLOW. 
%%%%%%%%%%%%%%%%%%%%%%%%%%%%%%%%%%%%%%%%%%%%%%%%%%%%%%%%%%%%%%%%%%%%%%%%%%%%%%%
% \subsection{Bibliography}
\bibliographystyle{abbrv}
\bibliography{semu}

\end{document}